\documentclass[conference]{IEEEtran}
\usepackage{cite}
\usepackage{graphicx}
\usepackage{amsmath}
\usepackage{bm}
\usepackage{url}

\begin{document}
\title{Channel Estimation for Intelligent Reflecting Surface Assisted Wireless Communications}
\author{\IEEEauthorblockN{Mangqing Guo and M. Cenk Gursoy}
\IEEEauthorblockA{Department of Electrical Engineering and Computer Science, Syracuse University, Syracuse, NY 13244.
\\
Email: mguo06@syr.edu, mcgursoy@syr.edu
}
}
\maketitle

\begin{abstract}
In this paper, the minimum mean square error (MMSE) channel estimation for intelligent reflecting surface (IRS) assisted wireless communication systems is investigated. In the considered setting, each row vector of the equivalent channel matrix from the base station (BS) to the users is shown to be Bessel $K$ distributed, and all these row vectors are independent of each other. By introducing a Gaussian scale mixture model, we obtain a closed-form expression for the MMSE estimate of the equivalent channel, and determine analytical upper and lower bounds on the mean square error. Using the central limit theorem, we conduct an asymptotic analysis of the MMSE estimate, and show that the upper bound on the mean square error of the MMSE estimate is equal to the asymptotic mean square error of the MMSE estimation when the number of reflecting elements at the IRS tends to infinity. Numerical simulations show that the gap between the upper and lower bounds are very small, and they almost overlap with each other at medium signal-to-noise ratio (SNR) levels and moderate number of elements at the IRS.
\end{abstract}


\section{Introduction}
The uncontrollability of channel conditions is one of the key challenges, limiting the efficiency of wireless communication systems. By introducing a large number of reconfigurable passive reflecting elements, the intelligent reflecting surface (IRS) could significantly improve the coverage and rate of wireless systems, and it has attracted much interest in recent years \cite{RenzoZapponeDebbahEtAl2020,BasarDiRenzoDeRosnyEtAl2019}.

For instance, the energy efficiency (EE) of IRS assisted wireless communication systems is studied in \cite{HuangZapponeAlexandropoulosEtAl2019}, where it is shown that the IRS-based resource allocation methods could improve EE by up to $300\%$, compared with the traditional multi-antenna amplify-and-forward relaying. A stochastic geometry analysis is performed in \cite{ZhuZhengWong2020} for large intelligent surface (LIS) assisted millimeter wave networks, and it is demonstrated that the LISs dramatically improve the average rate and area spectral efficiency of millimeter wave networks when the base station (BS) density is lower than the LIS density. In \cite{GuoLiangChenEtAl2019}, weighted sum-rate optimization has been performed by jointly optimizing the active beamforming at the BS and the passive beamforming at the IRS in IRS enhanced wireless networks.

Since there are no active components at the IRS to send or process pilot symbols for channel estimation, the acquisition of the channel state information (CSI) in IRS assisted wireless communication systems is different from that in traditional wireless networks \cite{NadeemAlwazaniKammounEtAl2020,ChenLiangChengEtAl2019,JensenDeCarvalho2020,LinWangFanEtAl2019,MirzaAli2020,HeYuan2020}. The channel estimation process in \cite{NadeemAlwazaniKammounEtAl2020,ChenLiangChengEtAl2019} is divided into several sub-frames, and the number of symbols transmitted in each sub-frame is greater than or equal to the number of users. In \cite{JensenDeCarvalho2020}, channel estimation is converted into an optimization problem, which can only be solved for a particular choice of the IRS phase quantization. A Lagrange multiplier and dual ascent based iterative algorithm for channel estimation in IRS assisted wireless communication systems is proposed in \cite{LinWangFanEtAl2019}. Compressed sensing algorithms are used in \cite{MirzaAli2020,HeYuan2020} to estimate the cascaded channel from the user to the BS through IRS.

In the prior work on channel estimation in IRS assisted wireless communication systems, proposed methods generally have high costs in terms of the computational complexity and number of pilot symbols. In this paper, we focus on finding an efficient channel estimation method for IRS assisted wireless communication systems. By considering the equivalent channel from the BS to the users, we can address the channel estimation problem employing conventional methods for wireless networks. With this approach, we determine that each row vector of the equivalent channel has a Bessel $K$ distribution, and all the rows are independent of each other. By introducing a Gaussian scale mixture model, we obtain the MMSE estimate of the equivalent channel, and identify analytical upper and lower bounds on the mean square error. Applying the central limit theorem, we perform an asymptotic analysis of the channel estimation, through which we show that the upper bound on the mean square error of the MMSE estimation equals the asymptotic mean square error of the MMSE estimation when the number of reflecting elements at the IRS tends to infinity. Compared with prior work on channel estimation in IRS assisted wireless communication systems, our proposed channel estimation method is completed in one stage via transmitting orthogonal pilots from the users, and enables us to obtain analytical expressions for the MMSE estimate of the equivalent channel coefficients through a more efficient scheme with low computational complexity.

The remainder of the paper is organized as follows. We describe the system model in Section \ref{sec:model} and analyze the statistics of the channel matrix in Section \ref{sec:statistics}. We derive the MMSE channel estimator in Section \ref{sec:mmse}. Numerical results are given in Section \ref{sec:numerical} and conclusions are drawn in Section \ref{sec:conclusion}.
 
\section{System Model} \label{sec:model}
\begin{figure}[htbp]
	\centering
	\includegraphics[width=3.5in]{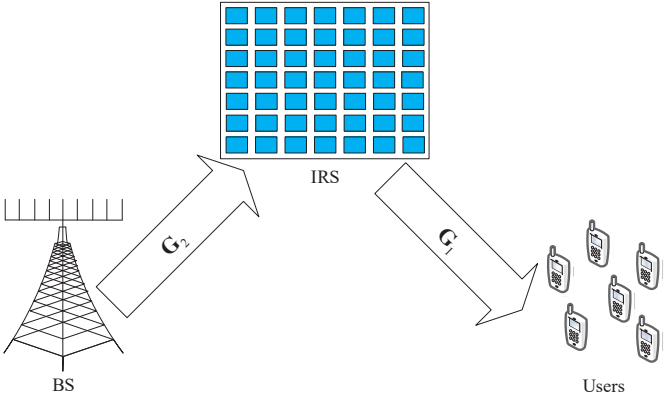}\\
	\caption{IRS assisted wireless communication systems.}\label{fig1}
\end{figure}

We consider an IRS assisted wireless communication system as depicted in Fig. \ref{fig1}, which is comprised of one BS equipped with $M$ antennas, an intelligent reflecting surface with $M_1$ reflecting elements, and $N$ users. All the users are uniformly distributed in a circle with radius $R$ and the IRS is located at the center of the circle. ${{\bf{G}}_1} \in {{\cal C}^{N \times {M_1}}}$, ${{\bf{G}}_2} \in {{\cal C}^{{M_1} \times M}}$ and ${\bf{V}} \in {{\cal C}^{{M_1} \times {M_1}}}$ denote the channel coefficients in the link from the IRS to the users, channel coefficients from the BS to the IRS, and the scattering matrix at the IRS, respectively. We assume that the direct link between the BS and users is not operational as a result of unfavorable propagation conditions (e.g., due to blockages) \cite{HuangZapponeAlexandropoulosEtAl2019,NadeemKammounChaabanEtAl2020}.

Furthermore, we assume flat fading channels between the BS, IRS and users, i.e., ${{\bf{G}}_1} = \text{diag}\left( {\sqrt {{{\bm{\beta }}_1}} } \right){{\bf{H}}_1}$ and ${{\bf{G}}_2} = \sqrt {{\beta _2}} {{\bf{H}}_2}$, where $\text{diag}\left( \cdot \right)$ is the diagonalization operation for a given vector \cite{KudathanthirigeGunasingheAmarasuriya2020,XiaShi2020}. ${{\bm{\beta }}_1} = [{\beta _{1,1}},{\beta _{1,2}}, \cdots ,{\beta _{1,N}}]$ is the path-loss vector from the IRS to the entire set of users, where ${\beta _{1,i}} = {\beta _{10}} + 10{\alpha _1}\log_{10} \left( {\frac{{{d_{1,i}}}}{{{d_{10}}}}} \right)$ is the path-loss from the IRS to the $i$th user in dB, where ${{d_{1,i}}}$ is the distance from the IRS to the $i$th user, ${{d_{10}}}$ is the reference distance for the path-loss between the IRS and users, ${\beta _{10}}$ is the corresponding path-loss at ${{d_{10}}}$, and $\alpha_1$ is the path-loss decay exponent. ${{\bf{H}}_1} \in {{\cal C}^{N \times {M_1}}}$ is the small-scale fading channel coefficients from the IRS to all the users, and all the elements of ${{\bf{H}}_1}$ are independent and identically distributed (i.i.d.) complex Gaussian random variables with zero mean and unit variance. Similarly, the path-loss between BS and IRS is ${\beta _2} = {\beta _{20}} + 10{\alpha _2}\log_{10} \left( {\frac{{{d_2}}}{{{d_{20}}}}} \right)$ in dB, where $d_2$ is the distance between the BS and IRS, $d_{20}$ is the reference distance for the path-loss between the BS and IRS, $\beta_{20}$ is the corresponding path-loss at $d_{20}$, and $\alpha_2$ is the path-loss decay exponent. All the elements of the small-scale fading matrix ${{\bf{H}}_2}$ are i.i.d. complex Gaussian random variables with zero mean and unit variance. Since path-loss changes slowly over time, it can be measured in advance and we assume ${{\bm{\beta }}_1}$ and ${\beta _2}$ are known at the BS.

As noted in the literature, we assume that the scattering matrix ${\bf{V}}$ is reconfigurable and known at the BS\cite{NadeemAlwazaniKammounEtAl2020,GongLuHoangEtAl2020}. We have ${\bf{V}} = \text{diag}\left( {{v_1}\exp (j{\theta _1}),{v_2}\exp (j{\theta _2}), \cdots ,{v_{{M_1}}}\exp (j{\theta _{{M_1}}})} \right)$, where ${v_i} \in [0,1]$ and ${\theta _i} \in [0,2\pi ]$ represent the amplitude and the phase coefficient for the $i$th element of the IRS, respectively. For the sake of simplicity in the analysis, we assume ${v_1} = {v_2} =  \cdots  = {v_{{M_1}}} = v$ in this paper.

\section{Statistics of the equivalent channel from the BS to the users} \label{sec:statistics}
In order to perform channel estimation at the BS, we need to know the distribution of the equivalent channel matrix from the BS to the users, which is denoted as ${\bf{G}} = {{\bf{G}}_1}{\bf{V}}{{\bf{G}}_2}$. We denote ${g_{ik}}$ as the element on the $i$th row and $k$th column of ${\bf{G}}$, and ${{\bf{g}}_i} \in {{\cal C}^{1 \times M}}$ as the $i$th row of ${\bf{G}}$. Additionally, ${h_{1,ik}}$, ${h_{2,ik}}$, ${{\bf{h}}_{1,i}}$ and ${{\bf{h}}_{2,i}}$ represent the elements on the $i$th row and $k$th column of ${{\bf{H}}_1}$ and ${{\bf{H}}_2}$, and the $i$th row of ${{\bf{H}}_1}$ and ${{\bf{H}}_2}$, respectively.
\subsection{Distribution of ${g_{ik}}$}
With the definitions given above, ${g_{ik}}$ can be expressed as
\begin{align}\label{equd1}
{g_{ik}} &= v\sqrt {{\beta _{1,i}}{\beta _2}} \sum\limits_{m = 1}^{{M_1}} {{h_{1,im}}{h_{2,mk}}\exp (j{\theta _m})} \nonumber\\
&= v\sqrt {{\beta _{1,i}}{\beta _2}} \sum\limits_{m = 1}^{{M_1}} {{h_{1,im}}{h_{3,mk}}}
\end{align}
where ${h_{3,mk}} = {h_{2,mk}}\exp (j{\theta _m})$, and ${h_{3,mk}} \sim {\cal C}{\cal N}(0,1)$ as a result of the distribution of ${h_{2,mk}}$. Since ${{h_{1,im}}}$ and ${{h_{2,mk}}}$ are independent of each other, ${{h_{1,im}}}$ is also independent of ${{h_{3,mk}}}$.

Let us denote ${g_{ik}} = {g_1} + j{g_2}$ and $t  = {t_1} + j{t_2}$. Then, the characteristic function of $g_{ik}$ is \cite{MallikSagias2011}
\begin{equation}\label{equd2}
{\Psi _{{g_{ik}}}}({t_1},{t_2}) = \frac{1}{{{{\left( {1 + \frac{{{\beta _{1,i}}{\beta _2}{v^2}}}{4}\left( {t_1^2 + t_2^2} \right)} \right)}^{{M_1}}}}}.
\end{equation}
Thus, the probability density function of ${g_{ik}}$ can be obtained as
\begin{align}\label{equd3}
&{p_{{g_{ik}}}}({g_1},{g_2}) \nonumber\\
=& \frac{1}{{4{\pi ^2}}}\int_{ - \infty }^\infty  {\int_{ - \infty }^\infty  {{\Psi _{{g_{ik}}}}({t_1},{t_2})\exp \left( { - j\left( {{g_1}{t_1} + {g_2}{t_2}} \right)} \right)d{t_1}} d{t_2}}   \nonumber\\
=& \frac{1}{{4{\pi ^2}}}\int_{ - \infty }^\infty  {\int_{ - \infty }^\infty  {\frac{{\exp \left( { - j\left( {{g_1}{t_1} + {g_2}{t_2}} \right)} \right)}}{{{{\left( {1 + \frac{{{\beta _{1,i}}{\beta _2}{v^2}}}{4}\left( {t_1^2 + t_2^2} \right)} \right)}^{{M_1}}}}}d{t_1}} d{t_2}}   \nonumber\\
=& \frac{1}{{4{\pi ^2}}}\int_0^\infty  {\int_0^{2\pi } {\frac{{r\exp \left( { - jr\left( {{g_1}\cos (\theta ) + {g_2}\sin (\theta )} \right)} \right)}}{{{{\left( {1 + \frac{{{\beta _{1,i}}{\beta _2}{v^2}{r^2}}}{4}} \right)}^{{M_1}}}}}d\theta } dr}  \nonumber\\
=& \frac{1}{{2\pi }}\int_0^\infty  {\frac{r}{{{{\left( {1 + \frac{{{\beta _{1,i}}{\beta _2}{v^2}{r^2}}}{4}} \right)}^{{M_1}}}}}{J_0}\left( {r\sqrt {g_1^2 + g_2^2} } \right)dr} \nonumber\\
=& \frac{{2{{\left( {g_1^2 + g_2^2} \right)}^{\frac{{{M_1} - 1}}{2}}}}}{{\pi \Gamma ({M_1}){{\left( {{\beta _{1,i}}{\beta _2}{v^2}} \right)}^{\frac{{{M_1} + 1}}{2}}}}}{K_{{M_1} - 1}}\left( {\frac{2}{{v\sqrt {{\beta _{1,i}}{\beta _2}} }}\sqrt {g_1^2 + g_2^2} } \right)
\end{align}
where ${J_0}( \cdot )$ represents zeroth-order Bessel function of the first kind, and ${K_n}( \cdot )$ denotes $n$th-order modified Bessel function of the second kind.
\subsection{Distribution of ${{\bf{g}}_i}$}
From (\ref{equd1}), we can obtain that
\begin{equation}\label{equd4}
{g_{i{j_1}}}{g_{i{j_2}}} = {\beta _{1,i}}{\beta _2}{v^2}\sum\limits_{{m_1} = 1}^{{M_1}} {\sum\limits_{{m_2} = 1}^{{M_1}} {{h_{1,i{m_1}}}{h_{1,i{m_2}}}{h_{3,{m_1}{k_1}}}{h_{3,{m_2}{k_2}}}} }.
\end{equation}
Therefore, $E\left\{ {{g_{i{j_1}}}{g_{i{j_2}}}} \right\} = 0$, where $E\{  \cdot \} $ stands for the expectation of the random variable. As $E\{ {g_{i{j_1}}}\}  = E\{ {g_{i{j_2}}}\}  = 0$, we have $\textit{cov}\left( {{g_{i{j_1}}},{g_{i{j_2}}}} \right) = E\left\{ {{g_{i{j_1}}}{g_{i{j_2}}}} \right\} = 0$, where $\textit{cov}(a,b)$ represents the covariance between random variables $a$ and $b$. Therefore, all the elements of ${{\bf{g}}_i}$ are uncorrelated with each other. Moreover, $p({g_{ik}}|{{\bf{h}}_{1,i}}) \sim {\cal C}{\cal N}\left( {0,{\beta _{1,i}}{\beta _2}{v^2}\sum\limits_{m = 1}^{{M_1}} {{{\left| {{h_{1,im}}} \right|}^2}} } \right)$. Thus, $p({{\bf{g}}_i}|{{\bf{h}}_{1,i}}) \sim {\cal C}{\cal N}\left( {0,\left( {{\beta _{1,i}}{\beta _2}{v^2}\sum\limits_{m = 1}^{{M_1}} {{{\left| {{h_{1,im}}} \right|}^2}} } \right){{\bf{I}}_M}} \right)$, where ${{\bf{I}}_M}$ denotes the identity matrix with dimension $M$. Then, we can obtain the joint probability density function of ${{\bf{g}}_i}$ and ${{\bf{h}}_{1,i}}$ as
\small
\begin{align}\label{equd5}
& p({{\bf{g}}_i},{{\bf{h}}_{1,i}}) \nonumber\\
=& p({{\bf{g}}_i}|{{\bf{h}}_{1,i}})p({{\bf{h}}_{1,i}})  \nonumber\\
=& \frac{1}{{{\pi ^{M + {M_1}}}{{\left( {{\beta _{1,i}}{\beta _2}{v^2}\sum\limits_{m = 1}^{{M_1}} {{{\left| {{h_{1,im}}} \right|}^2}} } \right)}^M}}}\exp \left( {\varphi ({\beta _{1,i}},{\beta _2},v,{{\bf{g}}_i},{{\bf{h}}_{1,i}})} \right)
\end{align}
\normalsize
where
\begin{equation}\label{equd6}
\varphi ({\beta _{1,i}},{\beta _2},v,{{\bf{g}}_i},{{\bf{h}}_{1,i}}) =  - \frac{{{{\left\| {{{\bf{g}}_i}} \right\|}^2}}}{{{\beta _{1,i}}{\beta _2}{v^2}\sum\limits_{m = 1}^{{M_1}} {{{\left| {{h_{1,im}}} \right|}^2}} }} - {\left\| {{{\bf{h}}_{1,i}}} \right\|^2}.
\end{equation}
Now, the probability density function of ${{\bf{g}}_i}$ can be obtained as
\begin{align}\label{equd7}
&p_{\bf{g}}({{\bf{g}}_i}) \nonumber\\
=& \int_{{C^{{M_1}}}} {p({{\bf{g}}_i},{{\bf{h}}_{1,i}})d{{\bf{h}}_{1,i}}}   \nonumber\\
=& \frac{2}{{{\pi ^M}\Gamma ({M_1}){{\left( {{\beta _{1,i}}{\beta _2}{v^2}} \right)}^M}}}* \nonumber\\
&{\kern 40pt}\int_0^\infty  {{r^{2{M_1} - 2M - 1}}\exp \left( { - \frac{{{{\left\| {{{\bf{g}}_i}} \right\|}^2}}}{{{\beta _{1,i}}{\beta _2}{v^2}{r^2}}} - {r^2}} \right)dr}   \nonumber\\
=& \frac{{2{{\left\| {{{\bf{g}}_i}} \right\|}^{{M_1}-M}}}}{{{\pi ^M}\Gamma ({M_1}){{\left( {{\beta _{1,i}}{\beta _2}{v^2}} \right)}^{\frac{{M + {M_1}}}{2}}}}}{K_{{M_1} - M}}\left( {\frac{2}{{v\sqrt {{\beta _{1,i}}{\beta _2}} }}\left\| {{{\bf{g}}_i}} \right\|} \right).
\end{align}

\subsection{Asymptotic distribution of ${{\bf{g}}_i}$ as ${M_1} \to \infty $}
Let us denote ${h_{imk}} = {h_{1,im}}{h_{3,mk}}$. From (\ref{equd3}), we can obtain that $p({h_{imk}}) = \frac{2}{\pi }{K_0}\left( {2\left| {{h_{imk}}} \right|} \right)$. Since ${h_{1,im}}$ and ${h_{3,mk}}$ are both complex Gaussian random variables with zero mean and unit variance and they are independent of each other, the components of the summation in (\ref{equd1}) are independent and identically distributed with zero mean and variance ${\beta _{1,i}}{\beta _2}{v^2}$. Therefore, according to central limit theorem, ${g_{ik}} \sim {\cal C}{\cal N}(0,{M_1}{\beta _{1,i}}{\beta _2}{v^2})$ as ${M_1}$ grows. As noted above, all the elements of ${{{\bf{g}}_i}}$ are uncorrelated with each other. Thus, we approximately have ${{\bf{g}}_i} \sim {\cal C}{\cal N}(0,{M_1}{\beta _{1,i}}{\beta _2}{v^2}{{\bf{I}}_M})$ for sufficiently large ${M_1}$.

\section{MMSE channel estimation} \label{sec:mmse}
We assume that the channel estimation is performed at the BS, and $N$ symbols are used for pilot transmission during each channel coherence interval. We denote the pilot matrix as ${\bf{P}}$, and assume that ${\bf{P}}$ is a unitary matrix. The received pilot signal at the BS can be expressed as
\begin{equation}\label{equ1}
	{\bf{Y}} = {\bf{PG}} + {\bf{\Phi }}
\end{equation}
where $\bf{\Phi }$ denotes the complex additive white Gaussian noise at the BS whose elements are independent and identically distributed with zero mean and variance ${\sigma ^2}$. Then, multiplying both sides of (\ref{equ1}) with ${{\bf{P}}^H}$, we can obtain
\begin{align}\label{equ2}	
\widetilde {\bf{Y}} &= {{\bf{P}}^H}{\bf{Y}} \nonumber\\
& = {\bf{G}} + {{\bf{P}}^H}{\bf{\Phi }}.
\end{align}
Since ${\bf{P}}$ is a unitary matrix, $\widetilde {\bf{\Phi }} = {{\bf{P}}^H}{\bf{\Phi }}$ has the same Gaussian distribution as ${\bf{\Phi }}$ with zero mean and variance ${\sigma ^2}$.

We have derived the probability density function for each row of ${\bf{G}}$ in the previous section, and it is obvious that different rows of ${\bf{G}}$ are independent of each other. Besides, the distribution of ${{\bf{g}}_i}$ can be equivalently represented as a Gaussian scale mixture \cite{KhazronSelesnick2008}
\begin{equation}\label{equ3}
{{\bf{g}}_i} = av\sqrt {{\beta _{1,i}}{\beta _2}} {{\bf{x}}_i}
\end{equation}
where ${{\bf{x}}_i} \sim {\cal C}{\cal N}(0,{{\bf{I}}_M})$, and $a$ is a scalar gamma random variable whose probability density function can be expressed as
\begin{equation}\label{equ4}
{p_A}(a) = \frac{{2{a^{2{M_1} - 1}}}}{{\Gamma ({M_1})}}\exp ( - {a^2}), {\kern 10pt} a > 0.
\end{equation}
Then, we can obtain
\begin{equation}\label{equ5}
\widetilde {\bf{Y}} = av{\kern 2pt}{\text{diag}}\left( {\sqrt {{{\bm{\beta }}_1}{\beta _2}} } \right){\bf{X}} + \widetilde {\bf{\Phi }}
\end{equation}
where ${\bf{X}} = {\left[ {{\bf{x}}_1^T,{\bf{x}}_2^T, \cdots ,{\bf{x}}_N^T} \right]^T}$, and ${{\bf{x}}_i} \sim {\cal C}{\cal N}(0,{{\bf{I}}_M})$. For a given $a$, the two components of $\widetilde {\bf{Y}}$, $av{\kern 2pt}{\text{diag}}\left( {\sqrt {{{\bf{\beta }}_1}{\beta _2}} } \right){\bf{X}}$ and $\widetilde {\bf{\Phi }}$, are both complex Gaussian random variables. Therefore, the conditional MMSE estimate of ${\bf{G}}$ is
\begin{equation}\label{equ6}
\widehat {\bf{G}}(a) = {\text{diag}}\left( {\frac{{{{\bm{\beta }}_1}{\beta _2}{a^2}{v^2}}}{{{{\bm{\beta }}_1}{\beta _2}{a^2}{v^2} + {\sigma ^2}}}} \right)\widetilde {\bf{Y}}.
\end{equation}
Then, we can obtain the MMSE estimate of ${\bf{G}}$ as in (\ref{equ8}) given at the top of next page, 
\begin{figure*}
\begin{align}\label{equ8}
\widehat {\bf{G}} = \int_0^\infty  {\widehat {\bf{G}}(a){p_A}(a)da} = {M_1}{\kern 2pt}{\text{diag}}\left( {{{\left( {\frac{{{\sigma ^2}}}{{{{\bm{\beta }}_1}{\beta _2}{v^2}}}} \right)}^{{M_1}}}{\Gamma _1}\left( { - {M_1},\frac{{{\sigma ^2}}}{{{{\bm{\beta }}_1}{\beta _2}{v^2}}}} \right)\exp \left( {\frac{{{\sigma ^2}}}{{{{\bm{\beta }}_1}{\beta _2}{v^2}}}} \right)} \right)\widetilde {\bf{Y}}
\end{align}
\hrulefill
\end{figure*}
where ${\Gamma _1}\left( {a,z} \right) = \int_z^\infty  {{t^{a - 1}}\exp ( - t)dt} $ is the upper incomplete gamma function.

Now, let us consider the mean square error of the MMSE estimate $\widehat {\bf{G}}$, which we denote as ${\text{mse}}\left( {\widehat {\bf{G}}} \right)$. For brevity in the description, we denote the $i$th row of $\widehat {\bf{G}}$ as ${{{\widehat {\bf{g}}}_i}}$, and the corresponding mean square error as ${\text{mse}}\left( {{{\widehat {\bf{g}}}_i}} \right)$. From (\ref{equ3}), we know that given $a$, ${{\bf{g}}_i}$ is conditionally distributed as ${\cal C}{\cal N}(0,{\beta _{1,i}}{\beta _2}{a^2}{v^2}{{\bf{I}}_M})$, and the probability density function of $a$ is ${p_A}(a)$. In this case, ${\text{mse}}\left( {\widehat {\bf{G}}} \right)$ has no simple analytical expression, and we can only obtain the following upper and lower bounds \cite{FlamChatterjeeKansanenEtAl2012}:
\begin{align}\label{equ9}
& \int_0^\infty  {\frac{{{\beta _{1,i}}{\beta _2}{a^2}{v^2}{\sigma ^2}}}{{{\beta _{1,i}}{\beta _2}{a^2}{v^2} + {\sigma ^2}}}{p_A}(a)da}  \leq {\text{mse}}\left( {{{\widehat {\bf{g}}}_i}} \right)  \nonumber\\
& {\kern 80pt} \leq \frac{{{\beta _{1,i}}{\beta _2}{v^2}{\sigma ^2}\int_0^\infty  {{a^2}{p_A}(a)da} }}{{{\beta _{1,i}}{\beta _2}{v^2}\int_0^\infty  {{a^2}{p_A}(a)da}  + {\sigma ^2}}}.
\end{align}
Substituting (\ref{equ4}) into (\ref{equ9}), we can further obtain
\begin{align}\label{equ10}
&{M_1}\frac{{{\sigma ^{2({M_1} + 1)}}}}{{{{\left( {{\beta _{1,i}}{\beta _2}{v^2}} \right)}^{{M_1}}}}}{\Gamma _1}\left( { - {M_1},\frac{{{\sigma ^2}}}{{{\beta _{1,i}}{\beta _2}{v^2}}}} \right)\exp \left( {\frac{{{\sigma ^2}}}{{{\beta _{1,i}}{\beta _2}{v^2}}}} \right) \leq \nonumber\\
&{\kern 100pt} {\text{mse}}\left( {{{\widehat {\bf{g}}}_i}} \right) \leq \frac{{{M_1}{\beta _{1,i}}{\beta _2}{v^2}{\sigma ^2}}}{{{M_1}{\beta _{1,i}}{\beta _2}{v^2} + {\sigma ^2}}}.
\end{align}
Finally, we can determine the upper and lower bounds of ${\text{mse}}\left( {\widehat {\bf{G}}} \right)$ as
\small
\begin{align}\label{equ11}
& \hspace{-.3cm} \frac{1}{N}{\text{sum}}\left[ {{M_1}\frac{{{\sigma ^{2({M_1} + 1)}}}}{{{{\left( {{{\bm{\beta }}_1}{\beta _2}{v^2}} \right)}^{{M_1}}}}}{\Gamma _1}\left( { - {M_1},\frac{{{\sigma ^2}}}{{{{\bm{\beta }}_1}{\beta _2}{v^2}}}} \right)\exp \left( {\frac{{{\sigma ^2}}}{{{{\bm{\beta }}_1}{\beta _2}{v^2}}}} \right)} \right] \leq   \nonumber\\
&{\kern 60pt} {\text{mse}}\left( {\widehat {\bf{G}}} \right) \leq \frac{1}{N}{\text{sum}}\left[ {\frac{{{M_1}{{\bm{\beta }}_1}{\beta _2}{v^2}{\sigma ^2}}}{{{M_1}{{\bm{\beta }}_1}{\beta _2}{v^2} + {\sigma ^2}}}} \right]
\end{align}
\normalsize
where ${\text{sum}}\left[  \cdot  \right]$ represents the summation of all the elements in a vector.

In Section \uppercase\expandafter{\romannumeral3}.C, we have shown that ${{\bf{g}}_i} \sim {\cal C}{\cal N}(0,{M_1}{\beta _{1,i}}{\beta _2}{v^2}{{\bf{I}}_M})$ as ${M_1}$ gets larger. Therefore, the MMSE estimate of ${\bf{G}}$ for large $M_1$ is
\begin{equation}\label{equ12}
\widetilde {\bf{G}} = {\text{diag}}\left( {\frac{{{M_1}{{\bm{\beta }}_1}{\beta _2}{v^2}}}{{{M_1}{{\bm{\beta }}_1}{\beta _2}{v^2} + {\sigma ^2}}}} \right)\widetilde {\bf{Y}},
\end{equation}
and the correspond mean square error is
\begin{equation}\label{equ13}
{\text{mse}}\left( {\widetilde {\bf{G}}} \right) = \frac{1}{N}{\text{sum}}\left[ {\frac{{{M_1}{{\bm{\beta }}_1}{\beta _2}{v^2}{\sigma ^2}}}{{{M_1}{{\bm{\beta }}_1}{\beta _2}{v^2} + {\sigma ^2}}}} \right]
\end{equation}
which is equal to the upper bound of ${\text{mse}}\left( {\widehat {\bf{G}}} \right)$ in (\ref{equ11}).

\section{Numerical Analysis} \label{sec:numerical}
We assume that there are 20 users in the IRS assisted wireless communication system, and all the users are uniformly distributed in a circular region with radius 1000m. The minimum distance between the IRS and a user is 500m. There are 20 antennas equipped at the BS. The reference distance for the path-loss between the BS, IRS and users are ${d_{10}} = {d_{20}} = 1$m, and the correspond path-loss decay exponents are ${\alpha _1} = 2$ and ${\alpha _2} = 2.8$, respectively. The path-loss at the reference distance is 30 dB, and the distance from the BS to the IRS is ${d_2} = 100$m. The phase coefficients of the IRS reflecting elements are uniformly distributed in $[0,2\pi ]$.

\begin{figure}[htbp]
	\centering
	\includegraphics[width=3.5in]{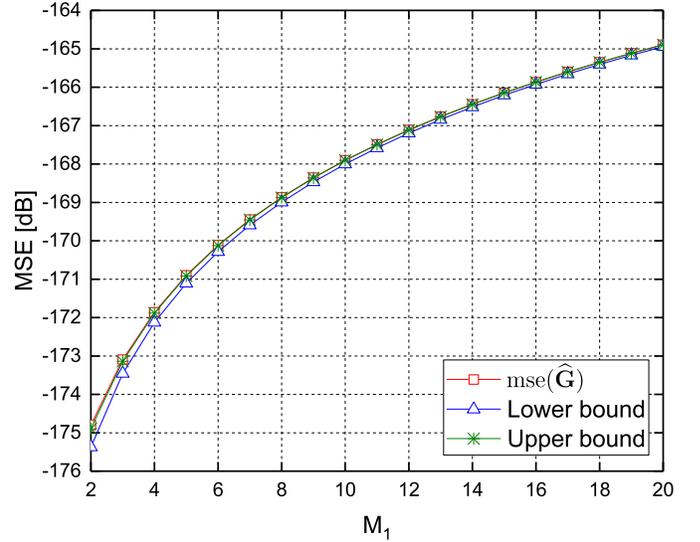}\\
	\caption{Mean square error of the MMSE estimate ${\text{mse}}(\widehat {\bf{G}})$, and the corresponding analytical upper and lower bounds versus the number of reflecting elements at IRS when SNR=0 dB and $v=1$.}\label{fig2}
\end{figure}

Fig. \ref{fig2} plots the mean square error of the MMSE estimate ${\text{mse}}(\widehat {\bf{G}})$, and the corresponding analytical upper and lower bounds versus the number of reflecting elements at the IRS when SNR$=0$ dB and $v=1$. ${\text{mse}}(\widehat {\bf{G}})$ represents the Monte Carlo simulation results of the mean square error for the MMSE estimate in (\ref{equ8}). The ``Upper bound'' and ``Lower bound'' denote the numerical results of the analytical bounds presented in (\ref{equ11}). These curves show that the gap between the upper and lower bound are very small. When $M_1$ is larger than 4, the upper bound overlaps with ${\text{mse}}(\widehat {\bf{G}})$, which matches well with our asymptotic analysis in (\ref{equ13}). We also note that since larger $M_1$ will introduce more uncertainties in the equivalent channel matrix ${\bf{G}}$, the mean square error increases as $M_1$ grows.

\begin{figure}[htbp]
	\centering
	\includegraphics[width=3.5in]{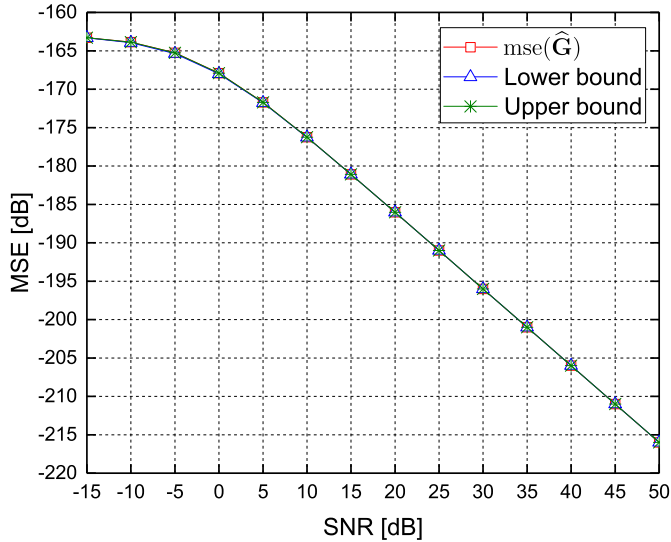}\\
	\caption{Mean square error of the MMSE estimate ${\text{mse}}(\widehat {\bf{G}})$, and the corresponding analytical upper and lower bounds versus SNR when the number of reflecting elements at IRS is 10 and $v=1$.}\label{fig3}
\end{figure}

Fig. \ref{fig3} plots the mean square error of the MMSE estimate ${\text{mse}}(\widehat {\bf{G}})$, and the corresponding analytical upper and lower bounds versus SNR when the number of reflecting elements at the IRS is 10 and $v=1$. These curves show that the mean square error decreases dramatically as SNR increases. In particular, mean square error in dB decays linearly with increasing SNR in dB when SNR is larger than 0 dB. We also observe that the upper and lower bounds, and ${\text{mse}}(\widehat {\bf{G}})$ again almost overlap with each other.

\begin{figure}[htbp]
	\centering
	\includegraphics[width=3.5in]{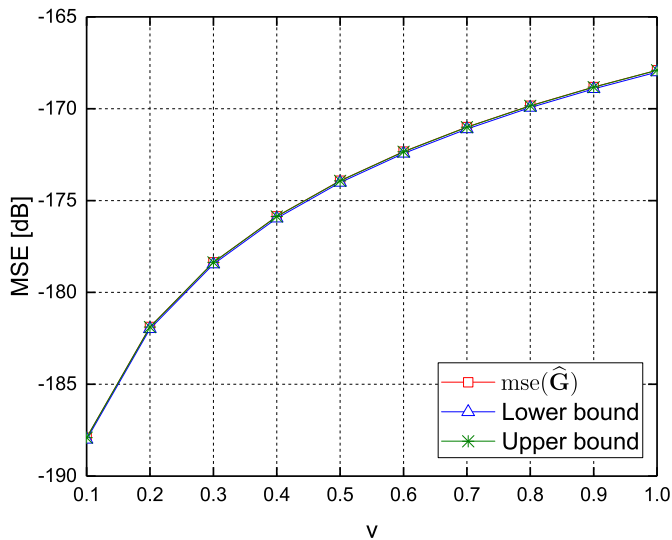}\\
	\caption{Mean square error of the MMSE estimate ${\text{mse}}(\widehat {\bf{G}})$, and the corresponding analytical upper and lower bounds versus the scattering amplitude at IRS when the number of reflecting elements at IRS is 10 and SNR=0 dB.}\label{fig4}
\end{figure}

Fig. \ref{fig4} plots the mean square error of the MMSE estimate ${\text{mse}}(\widehat {\bf{G}})$, and the corresponding analytical upper and lower bounds versus the scattering amplitude at IRS when the number of reflecting elements at the IRS is 10 and SNR$=0$ dB. These curves show that the mean square error increases as the scattering amplitude at the IRS grows, which could also be derived considering the monotonicity of the upper and lower bound expressions in (\ref{equ11}). The mean square error of the MMSE estimation overlaps with the upper bound, which further verifies our asymptotic characterization in (\ref{equ13}). As noted above, these curves also exhibit a very small gap between the upper and lower bounds.

\section{Conclusion} \label{sec:conclusion}
In this paper, we have analyzed the MMSE channel estimation in IRS assisted wireless communication systems. We have first identified statistics of the end-to-end channel matrix. Specifically, we have first shown that each row vector of the equivalent channel matrix from the BS to the users has a Bessel $K$ distribution, and all the rows are independent of each other. Following this characterization, we have employed a Gaussian scale mixture model, and obtained an analytical closed-form expression of the MMSE estimate of the equivalent channel. Furthermore, we have derived analytical upper and lower bounds of the mean square error. We have also provided an asymptotic analysis of the MMSE estimation, and shown that the upper bound of the mean square error of the MMSE estimate equals the asymptotic mean square error of the MMSE estimation as $M_1$ gets large.

We have further obtained certain characterizations. From the derived expressions of the MMSE estimate of the equivalent channel and the corresponding upper and lower bound expressions, we notice that the mean square error is independent of the number of antennas at the BS. Moreover, since the row vectors of the equivalent channel are independent of each other and orthogonal pilots are used during the channel estimation process, the mean square error does not depend on the number of users either.

Via numerical analysis, we have identified how the mean square error varies as a function of SNR, the number of elements and the scattering amplitudes at the IRS. We have also demonstrated that the upper and lower bounds lead to very accurate approximations of the mean square error.


\bibliographystyle{IEEEtran}
\bibliography{IRS_con.bbl}

\end{document}